\documentclass[twocolumn,pra,showpacs,preprintnumbers,amsmath,amssymb]{revtex4}
\usepackage{color}
\usepackage{bm, graphicx, amsmath}
\usepackage[mathscr]{eucal}

\begin{document}

\title{Lower bound on phase noise of classical states}
\author{Mark Hillery$^{1,2}$}
\affiliation{$^{1}$Department of Physics and Astronomy, Hunter College of the City University of New York, 695 Park Avenue, New York, NY 10065 USA \\ 
$^{2}$Physics Program, Graduate Center of the City University of New York, 365 Fifth Avenue, New York, NY 10016}

\begin{abstract}
An uncertainty relation for the number and phase of a single-mode field state is derived.  It is then used to find a lower bound on the phase noise of a classical state.  Any state that violates this condition is nonclassical.  An example of such a nonclassical state is presented.  Because a nonclassical state can have less phase noise than a classical state with the same average photon number, nonclassical states can play a role in the measurement of small phase shifts.
\end{abstract}

\maketitle

\section{Introduction}
It is well known that if the photon-number noise of a state is small enough, then the state is nonclassical.  In particular, a state is nonclassical if its photon statistics are sub-Poissoninan \cite{loudon1}.  No corresponding rule has been presented for the phase noise.  That is, is there a limit such that if the phase noise of a state falls below that limit, then the state must be nonclassical?  It is this question that will be studied here.

One of the difficulties associated with answering this question is the lack of a true phase operator \cite{susskind,nieto}.  One can define a phase distribution in terms of a positive-operator-valued measure, but because it is not a projection-valued measure there is no hermitian phase operator associated with it \cite{shapiro1}.  It is also possible to develop this measure by defining a phase operator on a finite-dimensional space and, once expectation values have been calculated, taking the limit as the dimension goes to infinity \cite{pegg}.  Another, independent, approach to the study of phase in quantum mechanics has been through phase-space techniques and the use of quasi-probability distribution functions \cite{schleich,tanas,knight}.  Finally, we should note, as has been pointed out by Bergou and Englert, the quantum-classical correspondence does not lead to a unique prescription for a phase operator or functions thereof \cite{bergou}.

Here we shall adopt the approach of Susskind and Glowgower, which is based on the polar decomposition of the annihilation operator \cite{susskind}.  This produces an operator that is a partial isometry, and this operator is taken to correspond to the classical quantity $e^{i\phi}$, where $\phi$ is the phase of the single-mode field.  This operator can then be used to define the phase noise of a state.  We shall derive a number-phase uncertainty relation based on this expression.  The uncertainty relation can then be used to find a lower limit on the phase noise for classical states.

\section{Uncertainty relation}

As was mentioned in the Introduction, Susskind and Glowgower were the first to suggest that rather than attempt to define a phase operator one should consider operators that correspond to the classical quantitities $e^{\pm i\phi}$, where $\phi$ is the phase of a single-mode classical field.  Let us now consider a single-mode quantum field whose number operator and number states we shall denote  by $N$ and $|n\rangle$, respectively.  The actions of the Susskind-Glowgower operators $E_{+}$ and $E_{-}$ are given by
\begin{equation}
E_{+}|n\rangle = |n+1\rangle \hspace{5mm} E_{-}|n\rangle = (1-\delta_{n,0})|n-1\rangle .
\end{equation}
The operator $E_{+}$ corresponds to the classical quantity $e^{-i\phi}$ and $E_{-}$ corresponds to $e^{i\phi}$.  By taking proper linear combinations of $E_{+}$ and $E_{-}$ we can define operators that correspond to $\cos\phi$ and $\sin\phi$.  In particular, if $C$ is the operator corresponding to $\cos\phi$ and $S$ is the operator corresponding to $\sin\phi$, then
\begin{equation}
C=\frac{1}{2}(E_{-}+E_{+}) \hspace{5mm} S=\frac{1}{2i}(E_{-}-E_{+}) .  
\end{equation}
As a measure of phase noise, we shall adopt the quantity
\begin{equation}
\left\langle (E_{-} - \langle E_{-}\rangle )(E_{+} - \langle E_{+}\rangle ) \right\rangle = 1-|\langle E_{-}\rangle |^{2} .
\end{equation}
For states with a well defined phase, $|\langle E_{-}\rangle |$ is close to one, and the above quantity is small.  For a state with a poorly defined phase, $|\langle E_{-}\rangle |$ is small, and $1-|\langle E_{-}\rangle |^{2}$ is close to one.  Classically, if the phase fluctuations are not too large, the quantity in the above equation corresponds to $(\Delta\phi )^{2}$.  

The operators $C$ and $S$ obey the uncertainty relations 
\begin{equation}
\label{CS}
(\Delta N)^{2}(\Delta C)^{2} \geq \frac{1}{4} |\langle S\rangle |^{2} \hspace{5mm} (\Delta N)^{2}(\Delta S)^{2} \geq \frac{1}{4} |\langle C\rangle |^{2} .
\end{equation}
These expressions can be added to give us an upper bound for $|\langle E_{-}\rangle |^{2}$ in terms of $(\Delta N)^{2}$.  We first note that
\begin{equation}
\label{C2S2}
|\langle E_{-}\rangle |^{2} = \langle C\rangle^{2} + \langle S\rangle^{2} ,
\end{equation}
and
\begin{equation}
\label{deltaCS}
(\Delta C)^{2} + (\Delta S)^{2} = 1-\frac{1}{2}\langle P_{0}\rangle - |\langle E_{-}\rangle |^{2} ,
\end{equation}
where $P_{0}=|0\rangle\langle 0|$. Adding the two inequalities in Eq.\ (\ref{CS}) and taking Eqs.\ (\ref{C2S2}) and (\ref{deltaCS}) into account, yields
\begin{equation}
\label{unc1}
\left[ (\Delta N)^{2} + \frac{1}{4} \right] \left( 1-|\langle E_{-}\rangle |^{2}\right) \geq \frac{1}{4} + \frac{1}{2} \langle P_{0}\rangle (\Delta N)^{2} .
\end{equation}
Because the last term on the right-hand side is positive, we can drop it giving us the simpler, but less precise, relation
\begin{equation}
\label{unc}
\left[ (\Delta N)^{2} + \frac{1}{4} \right] \left( 1-|\langle E_{-}\rangle |^{2}\right) \geq \frac{1}{4} .
\end{equation}
This is an uncertainty relation for number and phase noise, which holds for any single-mode quantum state.  Unlike the uncertainty relations in Eq.\ (\ref{CS}), the right-hand side is not state dependent.

We want to explore how good an uncertainty relation Eq.\ (\ref{unc}) is by applying it to several different states.  We shall consider a number state, a coherent state, and a truncated phase state.  For a number state, $\Delta N=0$ and $\langle E_{-}\rangle = 0$.  The left-hand side of Eq.\ (\ref{unc}) is then $1/4$, so number states satisfy this relation as an equality.  If we now consider a coherent state $|\alpha\rangle$, we find that $(\Delta N)^{2}=|\alpha |^{2}$, and to order $1/|\alpha |^{2}$, \cite{loudon2}
\begin{equation}
1-|\langle E_{-}\rangle |^{2} \cong \frac{1}{4|\alpha |^{2}} , 
\end {equation}
so that for large $|\alpha |$ a coherent state is approximately a minimum-uncertainty state.  For the truncated phase state
\begin{equation}
|\theta , n_{0}\rangle = \frac{1}{(n_{0}+1)^{1/2}} \sum_{n=0}^{n_{0}} e^{in\theta} |n\rangle ,
\end{equation}
we have that
\begin{equation}
(\Delta N)^{2} = \frac{n_{0}^{2}}{12} + \frac{n_{0}}{6} , \hspace{5mm} |\langle E_{-}\rangle |=\frac{n_{0}}{n_{0}+1} .
\end{equation}
A truncated phase state is, therefore, not a minimum uncertainty state for Eq.\ (\ref{unc}), because the left-hand side increases linearly with $n_{0}$ while the right-hand side is a constant.  We can improve the situation somewhat if we consider instead Eq.\ (\ref{unc1}).  In that case, the left-hand side is still $n_{0}/6$, but now the right-hand side is approximately equal to $n_{0}/24$ for large $n_{0}$.  While the truncated phase state is not a minimum uncertainty state for Eq.\ (\ref{unc1}), both sides of this uncertainty relation increase linearly with $n_{0}$.  Consequently, for a truncated phase state, Eq.\ (\ref{unc1}) provides a considerably better bound than does Eq.\ (\ref{unc}).

\section{Classical states}
Now let us restrict our attention to classical states.  We want to use our uncertainty relation to derive a condition that the phase noise of a classical state must satisfy.  We begin by noting that for a coherent state $|\alpha\rangle$, Eq.\ (\ref{unc}) implies that 
\begin{equation}
\label{3.1}
\frac{1}{\left[ 1+\frac{1}{2|\alpha |^{2}} \right]^{1/2}} \geq |\langle\alpha |E_{-}|\alpha\rangle |,
\end{equation}
because for a coherent state $(\Delta N)^{2}=\langle N\rangle = |\alpha |^{2}$.  In the interest of notational simplicity, we shall denote the left-hand side of Eq.\ (\ref{3.1}) by $f(\alpha )$.  The density matrix of a classical state is given by 
\begin{equation}
\rho = \int d^{2}\alpha\, P(\alpha ) |\alpha\rangle\langle \alpha | ,
\end{equation}
where $P(\alpha )$ is nonnegative definite and no more singular than a $\delta$ function.  This implies that $P(\alpha )d^{2}\alpha$ is a probability measure.  Eq.\ (\ref{3.1}) now allows us to deduce that
\begin{equation}
\label{3.3}
\int d^{2}\alpha\, P(\alpha ) f(\alpha ) \geq \left| \int d^{2}\alpha\, P(\alpha )  \langle\alpha |E_{-}|\alpha\rangle \right| = |\langle E_{-}\rangle | .
\end{equation}
Application of the Schwarz inequality gives us that
\begin{eqnarray}
\left| \int d^{2}\alpha\, P(\alpha ) f(\alpha ) \right|^{2}  & \leq & \left[ \int d^{2}\alpha\, P(\alpha ) \right] \left[\int d^{2}\alpha\, P(\alpha ) |f(\alpha )|^{2} \right] \nonumber \\
&\leq & \int d^{2}\alpha\, P(\alpha ) |f(\alpha )|^{2} ,
\end{eqnarray} 
which, in conjunction with Eq.\ (\ref{3.3}) yields
\begin{equation}
|\langle E_{-}\rangle |^{2} \leq \int d^{2}\alpha\, P(\alpha ) |f(\alpha )|^{2} .
\end{equation} 
Using this result to find a bound on the phase noise we have that
\begin{eqnarray}
\label{3.6}
1 - |\langle E_{-}\rangle |^{2} & \geq &  \int d^{2}\alpha\, P(\alpha )\left[ 1-1 |f(\alpha )|^{2}\right] \nonumber \\
& \geq & \int d^{2}\alpha P(\alpha ) \frac{1}{1 + (2|\alpha |)^{2}} .
\end{eqnarray}
By one more application of the Schwarz inequality it is possible to find a lower bound for the right-hand side of the above inequality that is a function of the mean photon number.  We have that 
\begin{equation}
\left[  \int d^{2}\alpha\, P(\alpha ) \frac{1}{1 + (2|\alpha |)^{2}} \right] \left[ \int d^{2}\alpha P(\alpha ) \left(1 + (2|\alpha |)^{2} \right) \right] \geq 1.
\end{equation}
Combining this with Eq.\ (\ref{3.6}) we obtain
\begin{equation}
\label{classical}
1 - |\langle E_{-}\rangle |^{2} \geq \frac{1}{4\langle N\rangle + 1} ,
\end{equation}
which is our condition on the phase noise of a classical state.  Any state that does not satisfy this inequality, that is any state for which
\begin{equation}
\label{nonclassical}
1 - |\langle E_{-}\rangle |^{2} < \frac{1}{4\langle N\rangle + 1} ,
\end{equation}
is nonclassical.

In order to show that there are states that satisfy Eq.\ (\ref{nonclassical}), consider the state
\begin{equation}
|\psi\rangle = \frac{2\sqrt{3}}{\sqrt{n_{0}(n_{0}^{2}+2)}} \left[ \sum_{n=0}^{n_{0}/2} n|n\rangle + \sum_{n=(n_{0}/2)+1}^{n_{0}} (n_{0}-n)|n\rangle \right] ,
\end{equation}
where we have assumed that $n_{0}$ is even.  The number distribution of this state is peaked around $n_{0}/2$ but is very broad.  This allows the phase to be very well defined, and we find for large $n_{0}$,
\begin{equation}
1 - |\langle E_{-}\rangle |^{2} \cong \frac{12}{n_{0}^{2}} .
\end{equation}
For large $n_{0}$ the expectation value of the photon number is approximately $n_{0}/2$, so that
\begin{equation}
\frac{1}{\langle\psi |N|\psi\rangle + 1} \cong \frac{1}{2n_{0}} .
\end{equation}
Therefore, for large $n_{0}$ the state $|\psi\rangle$ satisfies Eq.\ (\ref{nonclassical}) and is nonclassical.

\section{Conclusion}
As in the case of number fluctuations, phase fluctuations in classical states cannot be too small.  On the other hand, there are are nonclassical states whose fluctuations are below this limit.  In particular, we have seen from the last section that classical states must have phase noise, as measured by the quantity $1-|\langle E_{-}\rangle |^{2}$, that is of order $1/\langle N\rangle$ or greater.  By constructing an example, we have seen that there are states for which the phase noise goes as $1/\langle N\rangle^{2}$.  The use of such a state in an interferometer should permit the measurement of much smaller phase shifts than would be possible with a classical state with the same average photon number.  In fact, one class of nonclassical states, squeezed states, has already been extensively investigated, because of the ability of these states to improve interferometric measurements \cite{caves,shapiro2}.  The lower bound on the phase noise of classical states makes it clear that nonclassical states have a role to play in highly accurate phase measurements.

\acknowledgments
I would like to thank Prof. Janos Bergou for useful conversations.  This work was supported by the National Science Foundation under Grant No. PHY-9201912 and by a grant from the City University of New York under the PSC-CUNY Research Award Program.

\end{document}